\documentclass[11pt,onecolumn,english,aps,prola,superscriptaddress]{revtex4}
\usepackage{graphicx,psfrag,amsmath,amssymb,float}
\usepackage{color,times}
\usepackage{graphicx,natbib,ulem}

\newcommand{\be}{\begin{equation}}
\newcommand{\ee}{\end{equation}}
\newcommand{\bea}{\begin{eqnarray}}
\newcommand{\eea}{\end{eqnarray}}

\newcommand{\bc}{\begin{center}}
\newcommand{\ec}{\end{center}}
\usepackage[colorlinks,bookmarks=false,citecolor=blue,linkcolor=red,urlcolor=blue]{hyperref}
\begin{document}
{{
\title{Supplementary material for "Impurity-induced magnetic order in low dimensional spin gapped materials"}
\author{J. Bobroff}
\affiliation{Laboratoire de Physique des Solides,
Universit\'e Paris-Sud, UMR-8502 CNRS, 91405 Orsay, France}
\author{N. Laflorencie}
\affiliation{Laboratoire de Physique des Solides,
Universit\'e Paris-Sud, UMR-8502 CNRS, 91405 Orsay, France}
\author{L. K. Alexander}
\affiliation{Laboratoire de Physique des Solides,
Universit\'e Paris-Sud, UMR-8502 CNRS, 91405 Orsay, France}
\author{A. V. Mahajan}
\affiliation{Department of Physics, Indian Institute of Technology Bombay, Mumbai 400076, India}
\author{B. Koteswararao}
\affiliation{Department of Physics, Indian Institute of Technology Bombay, Mumbai 400076, India}
\author{P. Mendels}
\affiliation{Laboratoire de Physique des Solides,
Universit\'e Paris-Sud, UMR-8502 CNRS, 91405 Orsay, France}
%
\begin{abstract}
In this supplementary material, we investigate further the impurity-induced freezing mechanism in a doped system of 3D weakly coupled ladders resembling Bi(Cu$_{1-x}$Zn$_x$)$_2$ZnPO$_6$ using QMC.
\end{abstract}


\maketitle
\section{Introduction}
In the previous letter~\cite{Bobroff09}, we have argued that the collective freezing of effective moments having a 3D extension $V_{\xi}\sim\xi_x\xi_y\xi_z$ at $T> T_g$ is actually controlled by the exponentially decaying 3D coupling of the general form
\be
|J_{\rm 3D}^{\rm eff}({\vec{r}})|\simeq J_{\rm 3D}\exp\left(-\frac{x}{\xi_x}-\frac{y}{\xi_y}-\frac{z}{\xi_z}\right),
\label{eq:eff}
\ee
expected to occur for the wide class of spin gapped materials~\cite{Sigrist96,Imada97,Yasuda2002,Laflo03-04,Wessel2001,Vojta2006}.
The average coupling $J_{\rm avg}$ taken over all possible $J_{\rm 3D}^{\rm eff}({\vec{r}})$ does account for the broad distribution of effective interactions and is just given by
\be
J_{\rm avg}=\langle |J_{\rm 3D}^{\rm eff}({\vec{r}})|\rangle \simeq J_{\rm 3D}\frac{{\rm x}V_{\xi}}{1+{\rm x}V_{\xi}}
\label{eq:Javg}
\ee
where $V_{\xi}\sim\xi_x\xi_y\xi_z$ is the magnetic volume occupied by each induced moment. We propose that this average coupling governs the ordering, i.e.  ${\rm{T}}_{\rm g}\simeq J_{\rm avg}$.
\section{Microscopic model}
We want to check such an analysis against QMC simulations on a diluted 3D  model of weakly coupled ladders (schematized in Fig.~\ref{fig:lattice}) with the following parameters: $J_{\perp}/J=0.1$ and $J_{3D}/J=0.05$ which, using a value of  $J\simeq 100$ K corresponds to a spin gap $\Delta \simeq 35$ K and a transverse 3D coupling $J_{3D} \simeq 5$ K.
We then introduced non-magnetic impurities (open circles in Fig.~\ref{fig:lattice}) 
and performed large scale QMC simulations on 3D samples of sizes $L\times L\times L/2$,  with $L=16, 24, 32, 48$, down to temperature $T/J=0.01$. 
We also performed disorder averaging over a large number N of independent disordered samples ranging from $N=500$ for $L=16$ to $N=100$ for $L=48$.
\begin{figure}[!ht]
\begin{center}
\includegraphics[width=0.75\columnwidth,clip]{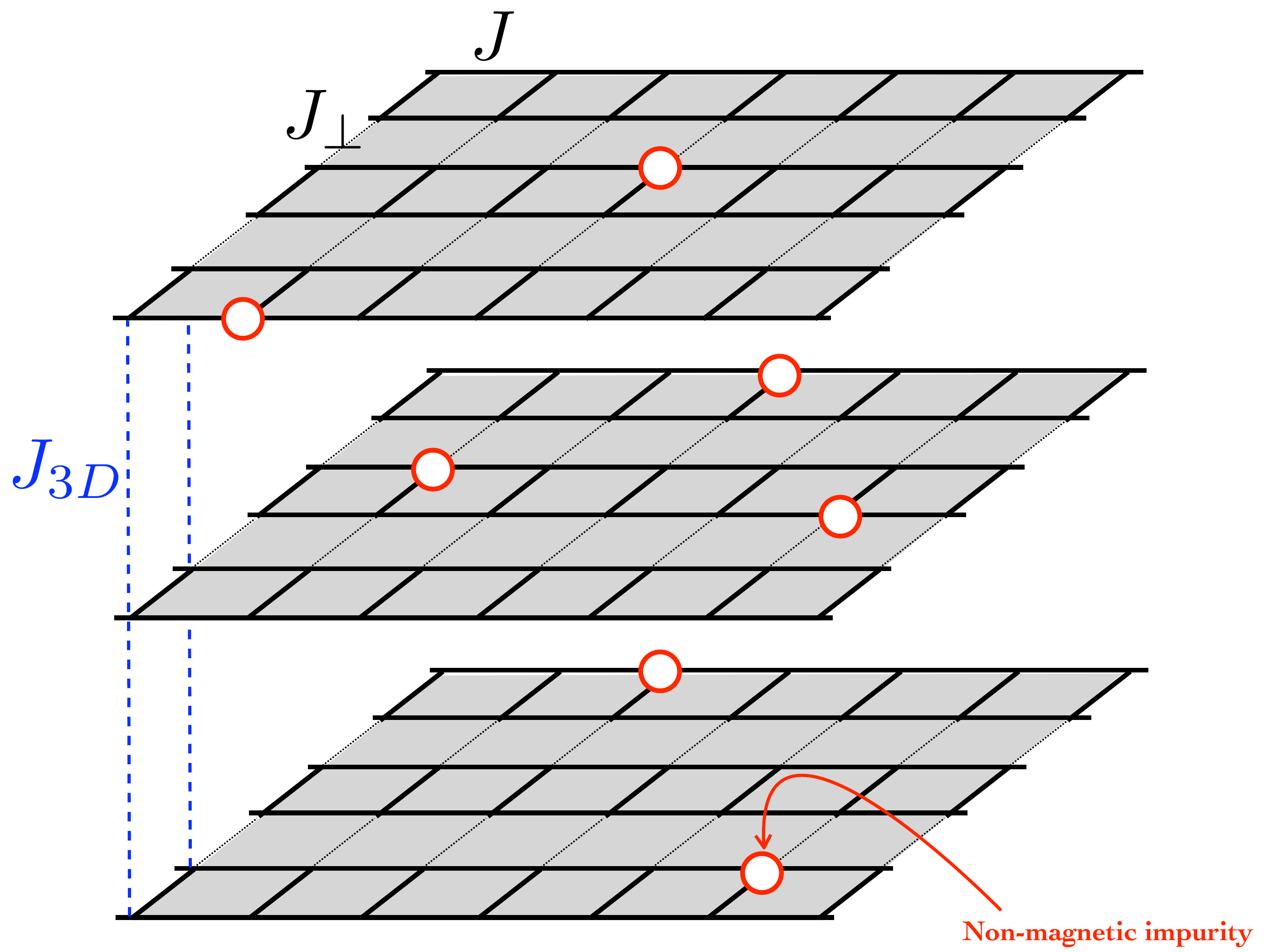}
\end{center}
\caption{(Color online) Schematic picture of the 3D system of coupled ladders used for the QMC simulations.}
\label{fig:lattice}
\end{figure}
\section{QMC results for the critical temperature}
The results for the 3D ordering temperature $T_g$ are shown in Fig.~\ref{fig:TG} versus the impurity concentration x. The transition was found by the standard technique using the finite size scaling of the spin stiffness at the transition point, as examplified in Fig.~\ref{fig:RHO} and discussed below. As shown in Fig.~\ref{fig:TG}, we get a linear increase for $T_g({\rm{x}})$ 
up to a threshold $\sim 3\%$ 
where $T_{g}$ saturates as expected. 
The linear part can be fitted by the form $T_g=2.8{\rm{x}}$ 
which compares quite well to our estimate Eq.~(\ref{eq:Javg}).
Indeed, with our parameters, we expect an average coupling $J_{\rm avg}\approx J_{3D}\times V_{\xi}\times {\rm{x}}\approx 0.05\times 30\times {\rm x}=1.5 {\rm x}$, meaning that with such a definition for $V_{\xi}$, we get  $T_g({\rm x})
\approx 2J_{\rm avg}$.

\begin{figure}[!ht]
\begin{center}
\includegraphics[width=0.75\columnwidth,clip]{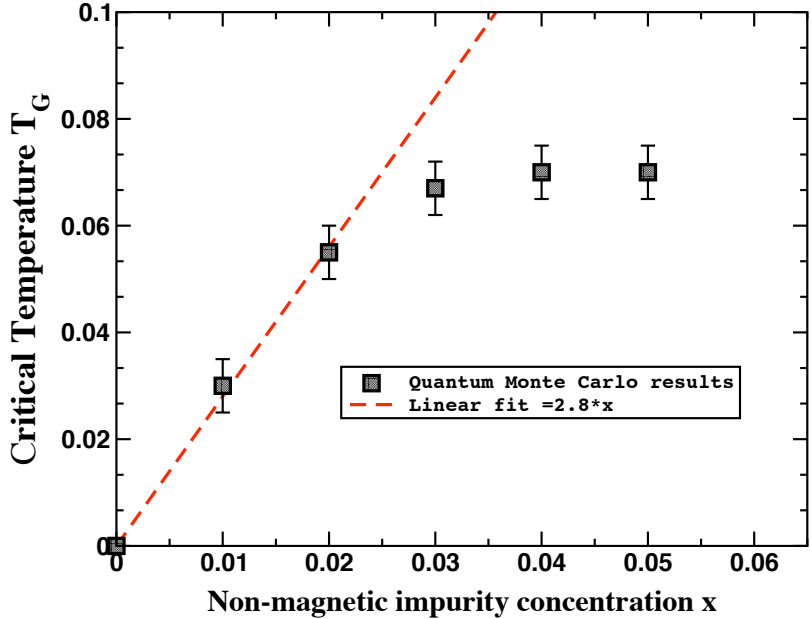}
\end{center}
\caption{(Color online) Critical temperature TG (in units of J) plotted versus impurity concentration x.}
\label{fig:TG}
\end{figure}

\section{Details about the critical point}
The way the critical ordering temperature was extracted from QMC simulations on finite size systems is actually standard since it relies on the finite size scaling of the order parameters, as for instance used in Ref.~\cite{Sandvik98}.
Therefore we computed the spin stiffness $\rho_s$, 
directly related to the square of the AF order parameter. In an 3D AF ordered phase $\rho_s$ 
is finite whereas it is 0 in a disordered phase.
At the critical point between the two regimes, there is a well-known finite size scaling 
\be
\rho_s(L)\sim L^{2-D-z},
\ee
where $D$ is the space dimension (here $D=3$) and $z$ is the dynamical exponent ($z=0$ for a finite temperature phase transition). 
Therefore we expect $\rho_s \times L$ 
to be a constant at the critical point where a crossing of the various system sizes should occur. 
We thus used such a criterion to identify the ordering transition at ${\rm{x}}=1,~2,~3,~4,~5\%$.
Results of such an analysis for a concentration ${\rm{x}}=2\%$ 
are displayed in Fig.~\ref{fig:RHO} 
for $L=16, 24, 32, 48$ 
with a critical point found at $T_{g}=0.055J$. 
In Fig.~\ref{fig:RHO}(A), we show the average stiffness versus $T/J$. 
In fact the spin stiffness is a directionnal quantity and can thus be computed in all space directions $x, y, z$, or averaged over all directions. The crossing of $\rho_s \times L$ 
is shown in Fig.~\ref{fig:RHO} (B) as well as in insets (X,Y,Z) for all the components of the stiffness. This clearly shows that the ordering is fully three dimensional and we find a remarkable agreement for the crossing temperatures in all directions at 
$T_g/J=0.055$.
\begin{figure}[!ht]
\begin{center}
\includegraphics[width=0.88\columnwidth,clip]{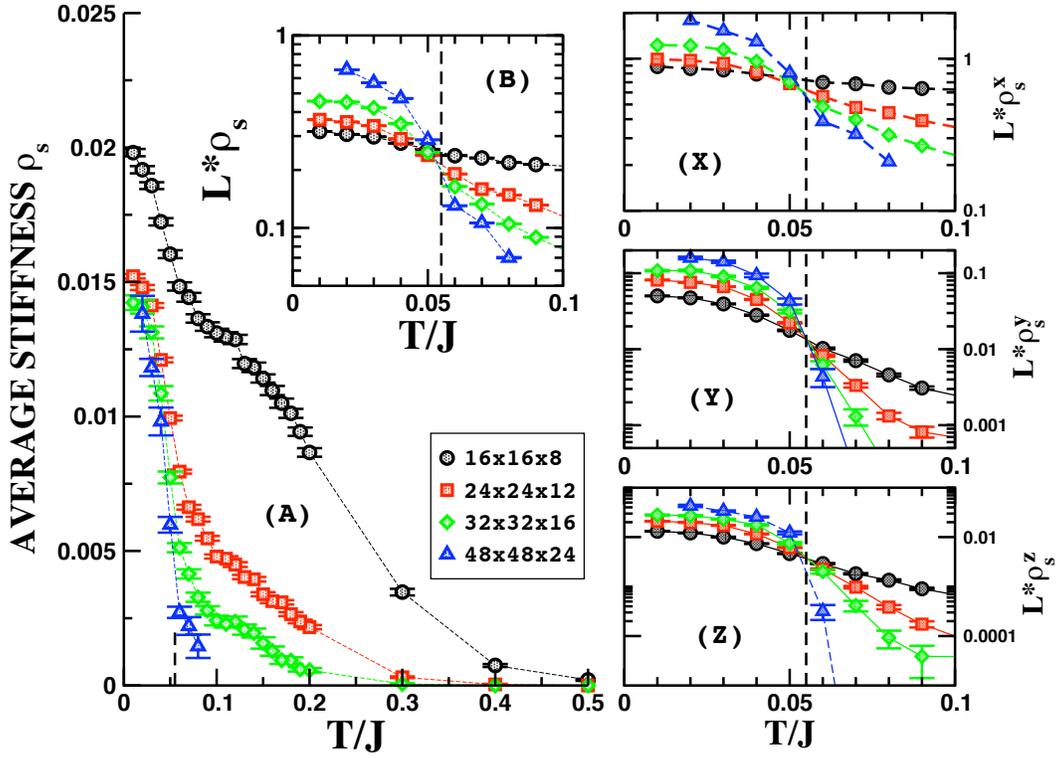}
\end{center}
\caption{(Color online) QMC results for the 3D model of coupled ladders with ${\rm x}=2\%$ of non-magnetic impurities. (A) shows the average 3D stiffness versus $T/J$. The same data are shown in (B) where $\rho_s\times L$ has a crossing for all sizes at $T/J=0.055$. Insets (X,Y,Z) show all the components of the spin stiffness times L that cross at the same critical temperature.}
\label{fig:RHO}
\end{figure}
\section{Comparison to experiments and conclusions}

To finally conclude on this issue of the 3D transition, we carefully checked that the ordering transition is a true AF 3D ordering which occurs at a freezing temperature $T_g$ proportionnal to the average coupling as proposed in the paper~\cite{Bobroff09}. We indeed confirm a linear regime with x at low concentration followed by a saturation a larger x corresponding to the fact that the average distance between impurity start to be of the order of the correlation length ${\xi}$. 
As a comparison we plotted on a common graph (Fig.~\ref{fig:TSTAR}) experimental results for $T_g$ rescaled to their ${\rm x}=3\%$ values for various spin-gapped materials together with the QMC results of this study. The agreement is very good.

\begin{figure}[!ht]
\begin{center}
\includegraphics[width=0.6\columnwidth,clip]{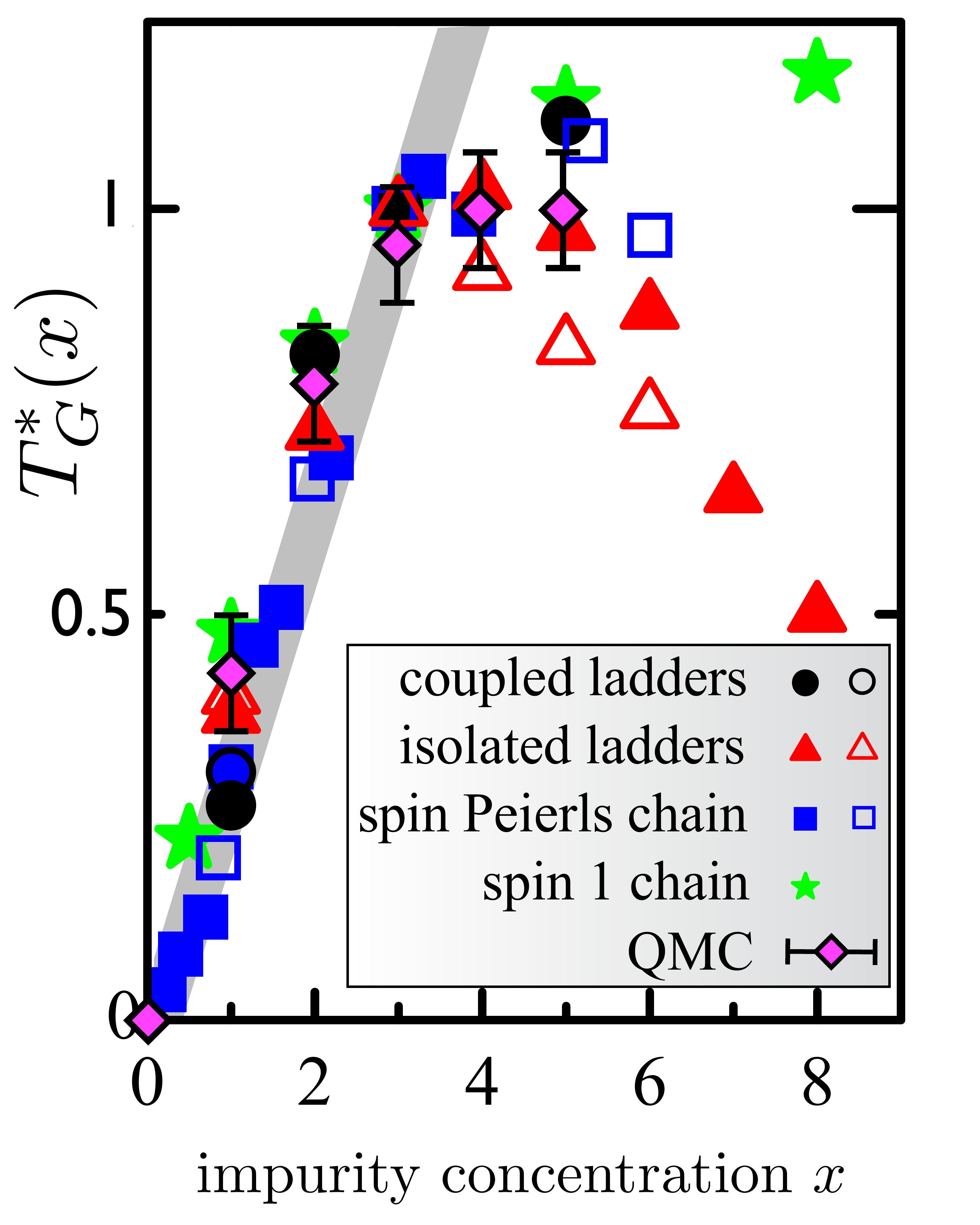}
\end{center}
\caption{(Color online) Transition temperatures $T^*_g$ (rescaled to their values at x=3$\%$) versus impurity concentration for various low-D spin-gapped systems: coupled ladders Bi(Cu$_{1-x}$(Zn or Ni)$_x$)$_2$PO$_6$ from this study; isolated ladder Sr(Cu$_{1-x}$(Zn or Ni)$_x$)$_2$O$_3$; Haldane chain Pb(Ni$_{1-x}$Mg$_x$)$_2$V$_2$O$_8$; spin-Peierls chains Cu$_{1-x}$(Zn or Ni)$_x$GeO$_3$. QMC data of Fig.~\ref{fig:TG} are also shown for comparison.}
\label{fig:TSTAR}
\end{figure}


\begin{thebibliography}{99}
\bibitem{Bobroff09} J. Bobroff, N. Laflorencie, L. K. Alexander, A. V. Mahajan, B. Koteswararao, and P. Mendels, Phys. Rev. Lett {\bf 103}, 047201 (2009).
\bibitem{Sigrist96} M. Sigrist and A. Furusaki, J. Phys. Soc. Jpn. {\bf 65}, 2385 (1996).
\bibitem{Imada97} M. Imada and Y. Iino, J. Phys. Soc. Jpn. {\bf 66}, 568 (1997).
\bibitem{Yasuda2002} C. Yasuda {\it et al.}, Phys. Rev. B {\bf 64}, 092405 (2001).
\bibitem{Laflo03-04} N. Laflorencie and D. Poilblanc, Phys. Rev. Lett. {\bf 90}, 157202 (2003).
\bibitem{Wessel2001} S. Wessel {\it et al.}, Phys. Rev. Lett. {\bf 86}, 1086 (2001).
\bibitem{Vojta2006} H. Weber and M. Vojta, Eur. Phys. J. B {\bf 53}, 185 (2006).
\bibitem{Sandvik98} A. W. Sandvik, Phys. Rev. Lett. {\bf 80}, 5196 (1998). 

\end{thebibliography}
\end{document}